\documentclass[12pt]{iopart}
\usepackage{epsfig}
\usepackage{iopams}
\begin{document}

\title[Probing the quantum coherence of a nanomechanical resonator II: Implementation]
{Probing the quantum coherence of a nanomechanical resonator using a superconducting qubit II: Implementation}

\author{M.P. Blencowe$^{\dag}$  and A.D. Armour$^{\ddag}$}
\address{\dag Department of Physics and Astronomy, 6127 Wilder
Laboratory, Dartmouth College, Hanover, NH 03755, USA\\
\ddag School of Physics and Astronomy, University of Nottingham,
Nottingham,\\ NG7 2RD, UK}
\ead{\mailto{miles.p.blencowe@dartmouth.edu},\\\mailto{andrew.armour@nottingham.ac.uk}}

\begin{abstract}
We describe a possible implementation of the nanomechanical quantum  
superposition generation and detection scheme described in the preceding, companion paper [Armour A D and Blencowe M P 2008 {\it New. J. Phys.} {\bf XX} XXX]. The  
implementation is based on the circuit quantum electrodynamics (QED) set-up, with the addition  
of a mechanical degree of freedom formed out of a suspended, doubly-clamped segment of the superconducting loop of a dc SQUID located directly opposite the centre conductor  
of a coplanar waveguide (CPW). The relative merits of two SQUID  
based qubit realizations  are addressed, in particular a
capacitively coupled charge qubit and inductively coupled flux qubit.  
It is found that both realizations are equally promising, with  
comparable qubit-mechanical resonator mode as well as qubit-microwave resonator mode  
coupling strengths.
  
\end{abstract}
\submitto{NJP}
\maketitle

\section{Introduction}

In reference~\cite{armournjp08}, we described a scheme for generating and  
detecting superposition states of a nanomechanical resonator. The  
scheme employs a qubit with dispersive coupling to the mechanical  
resonator. By preparing the qubit in a superposition of energy eigenstates and the  
mechanical resonator in a displaced thermal state, the latter  
subsequently evolves into a superposition state, the existence of  
which can be inferred through appropriate qubit state control and  
measurement.

\begin{figure}[t]
\centering \epsfig{file=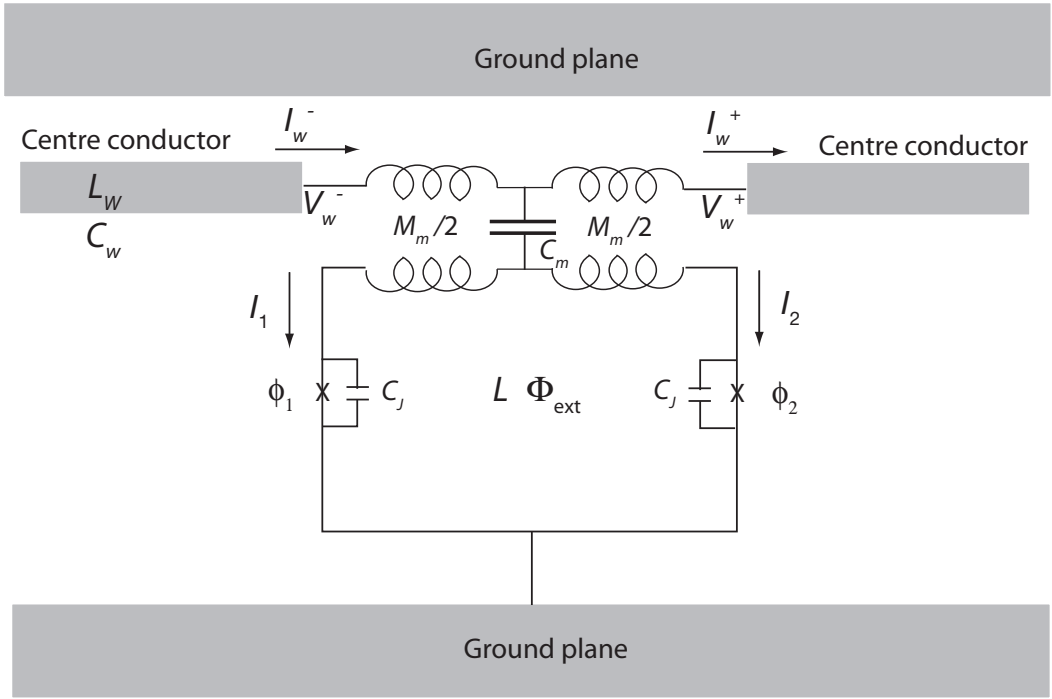,width=14cm} \caption{Model circuit of the closed CPW-dc SQUID system. The CPW comprises a uniform width centre conductor of length $l$ and parallel ground planes. The SQUID is located at the midway point  ($z=0$) in the gap between the centre conductor and one of the ground planes.  The mechanical resonator is formed out of a freely-suspended segment of the SQUID loop located directly opposite the centre conductor. An electromagnetic mode of the centre conductor couples both capacitively (via mutual capacitance $C_m$) and inductively (via mutual inductance $M_m$) to the SQUID loop circulating currents. Not shown are the  microwave qubit control/readout line and low frequency centre conductor bias line.} \label{schemefig}
\end{figure}

In the present paper, we describe a possible implementation of this  
scheme (see figure \ref{schemefig}) 
that is based on the demonstrated circuit quantum electrodynamics (QED) set-up~\cite{wallraffnature04,blaispra04}. 
Control and readout of the qubit is achieved by  
pumping and probing the appropriate mode of a microwave coplanar  
waveguide (CPW) resonator  that couples electromagnetically to the  
qubit. We shall consider two types of superconducting qubit: a  
capacitively coupled charge qubit, otherwise known as a Cooper pair  
box (CPB)~\cite{bouchiatps98,nakamuranature99}, and an inductively coupled flux qubit~\cite{chiorescunature04}. As we shall see,  
both types of qubit appear equally promising for implementation,  
having comparable strength couplings to the CPW resonator
and to the nanomechanical resonator. One motivation for adopting the  
circuit QED set-up is the adequate demonstrated qubit  
coherence times for our purpose~\cite{armournjp08}. 

We suppose that a  segment of the qubit flux loop directly opposite the centre conductor  
of the CPW is freely suspended, forming a doubly-clamped beam mechanical resonator. Related schemes can be found in references~\cite{buksprb06,zhouprl06,xuenjp07,blencoweprb07,bukspre07}.  
Furthermore, in addition to the usual  GHz microwave qubit control/readout line that is capacitively coupled to the CPW, we require a separate, low frequency (i.e., tens of MHz and below) centre conductor bias line, in order to achieve the necessary strong couplings between the  
mechanical resonator and both qubit and CPW mode, as well as to drive the mechanical  
resonator on resonance, producing  the displaced thermal state. In the case of the capacitively coupled charge qubit, 
the  required centre conductor voltage bias can be introduced via a series inductor,  
otherwise known as a `bias tee'  that is microwave engineered to  
reflect the separate,  much higher frequency GHz qubit control signal~\cite{regalpreprint08}. In the case of the inductively coupled flux qubit, the required current bias can be introduced via two superconducting low pass filters at each end of the centre conductor~\cite{arbelieee06}.  The  
high frequency thermal Johnson noise entering the bias line  can be significantly suppressed with additional low-pass filtering at each temperature stage of the set-up,  while at the same time allowing sufficiently large dc  voltage or current biases for the required mechanical resonator-qubit  
coupling strengths~\cite{schusterphd07}. On the other hand, the high frequency thermal  
 noise entering the microwave control line can be reduced by  
attenuating the signal~\cite{schusterphd07}; only low drive power signals are required to control  
the qubit~\cite{wallraffnature04}.

Sections 2 and 3 of this paper derive the respective charge and flux  
qubit Hamiltonians with coupled, single microwave and mechanical modes.  
Both qubit Hamiltonians have a common origin in the CPW-coupled-dc  
SQUID circuit of figure~\ref{schemefig}. Section 2 discusses the classical dynamics,  
while section 3 derives the quantum Hamiltonians of the charge and flux qubit  
systems, equations~(\ref{concisechargehamiltonianeq}) and (\ref{concisefluxhamiltonianeq}) respectively. While there is nothing particularly new in terms of theory in these two sections, they serve the purpose of mapping the scheme of reference~\cite{armournjp08} onto the circuit QED set-up with a unified treatment of the charge and flux qubit implementations, facilitating a direct comparison between them.   Readers not interested in the derivations may proceed  
directly to section 4, where the relative merits of the qubit  
Hamiltonians for actual device implementation are addressed. Section 5  
gives a brief conclusion. Appendix A  gives a derivation of the  
single mode approximation to the CPW superconducting phase wave  
equation, while appendix B derives the qubit Langevin equations.

\section{CPW-dc SQUID classical equations of motion }

In this section, we analyze the classical equations of motion for the phase field coordinate $\phi(z,t)$ of a length $l$ coplanar waveguide (CPW) with longitudinal coordinate $-l/2<z<l/2$ and gauge invariant phases $\phi_1$ and $\phi_2$ across each of the two Josephson junctions (JJ) of the dc SQUID.  These equations are the common starting point for the subsequent derivations of both the microwave resonator mode-flux qubit and microwave resonator mode-charge qubit Hamiltonians. Related analyses can be found in the review references~\cite{devoretpreprint04,wendinreview06,johanssonjpcm06,gellerreview06}. The CPW and SQUID are coupled via a mechanical, position-dependent mutual inductance $M_m$ and capacitance $C_m$ at $z=0$, arising from the adjacent, parallel centre conductor and suspended SQUID loop segments.  The position dependence and hence mechanical degree of freedom will be suppressed for the time being, to be  introduced in section~\ref{quantumequations} once the qubit Hamiltonians have been obtained. Figure~\ref{schemefig} details the closed CPW-SQUID circuit. Not shown are the  capacitively-coupled microwave signal pump-probe line, as well as the low frequency  bias line. Again, we will ignore these external lines, focusing first on the closed CPW-SQUID dynamics. The external bias dependences will be introduced later below once the relevant classical CPW-coupled SQUID Lagrangians have been derived [see equations(\ref{lagrangiancapeq}) and (\ref{lagrangianindeq})].   

For $-l/2<z<0$ and $0<z<l/2$, the CPW phase field coordinate obeys the wave equation
\begin{equation}
\frac{\partial^2\phi}{\partial t^2}=v^2 \frac{\partial^2\phi}{\partial z^2},
\label{waveeq}
\end{equation}
where $v=(L_w C_w)^{-1/2}$ is the phase velocity and $L_w$ and $C_w$ are the inductance and capacitance per unit length of the CPW, respectively. At the CPW boundaries $z=\pm l/2$, the current $I_w=-\frac{\Phi_0}{2\pi L_w}\frac{\partial\phi}{\partial z}$ vanishes, since as mentioned above we are in the first instance considering the CPW-SQUID system to be closed. We assume small CPW-SQUID couplings, i.e., $C_m/(C_w l)\ll 1$ and $M_m/(L_w l)\ll 1$. 

The equations of motion for the coupled SQUID can be conveniently derived by applying Kirchhoff's  laws for the voltages and currents at $z=0^{\pm}$ and in the SQUID loop. Referring to figure~\ref{schemefig}, current conservation gives
\begin{equation}
I_w^- -I_w^+ =I_1+I_2.
\label{currentconseq}
\end{equation}
The voltage relations are 
\begin{equation}
V_w^-=-\frac{M_m}{2}\frac{d I_1}{dt}+\frac{Q_m}{C_m}-\frac{M_m}{2}\frac{d I_w^-}{dt}+\frac{L}{2}\frac{d I_1}{dt}+\frac{\Phi_0}{2\pi}\frac{d\phi_1}{dt}\label{voltagerel1eq}
\end{equation}
and
\begin{equation}
V_w^+=-\frac{M_m}{2}\frac{d I_2}{dt}+\frac{Q_m}{C_m}+\frac{M_m}{2}\frac{d I_w^+}{dt}+\frac{L}{2}\frac{d I_2}{dt}+\frac{\Phi_0}{2\pi}\frac{d\phi_2}{dt},
\label{voltagerel2eq}
\end{equation}
where $V_w=\frac{\Phi_0}{2\pi}\frac{\partial\phi}{\partial t}$, $Q_m$ is the charge on the capacitance $C_m$ and $L$ is the SQUID loop self inductance.
Current conservation at the JJ nodes gives
\begin{eqnarray}
I_1&=&I_c \sin{\phi_1}+\frac{C_J\Phi_0}{2\pi}\ddot{\phi}_1\cr
I_2&=&I_c \sin{\phi_2}+\frac{C_J\Phi_0}{2\pi}\ddot{\phi}_2,
\label{currentJJreleq}
\end{eqnarray}
where $I_c$ is the JJ critical current and $C_J$ the JJ capacitance (assumed identical for each junction).
The JJ phases are related to the net flux $\Phi$ threading the SQUID loop as
\begin{equation}
\phi_2 -\phi_1=2\pi n +2 \pi\frac{\Phi}{\Phi_0},
\label{phasefluxeq}
\end{equation}
where the net flux is
\begin{equation}
\Phi=\Phi_{\mathrm{ext}}-\frac{M_m}{2}(I_w^- +I_w^+)+\frac{L}{2}(I_1 -I_2),
\label{fluxeq}
\end{equation}
with $\Phi_{\mathrm{ext}}$ the externally applied flux.  
It is more convenient to work with the SQUID phase variables $\gamma_{\pm}=(\phi_1 \pm\phi_2)/2$, in which case equation (\ref{currentJJreleq}) becomes
\begin{equation}
\omega_J^{-2}\ddot{\gamma}_+ +\sin\gamma_+ \cos\gamma_- =\frac{I_1+I_2}{2 I_c}=\frac{I_w^- -I_w^+}{2 I_c}
\label{gammapluseq}
\end{equation}
and
\begin{equation}
\omega_J^{-2}\ddot{\gamma}_- +\cos\gamma_+ \sin\gamma_- =\frac{M_m}{L}\frac{I^-_w +I^+_w}{2 I_c} -\frac{2}{\beta_L}\left[\gamma_-  +\pi \left(n+\frac{\Phi_{\mathrm{ext}}}{\Phi_0}\right)\right],
\label{gammaminuseq}
\end{equation}
where we have used equations (\ref{phasefluxeq}) and (\ref{fluxeq}), $\omega_J=\sqrt{2\pi I_c/(C_J \Phi_0)}$ is the JJ plasma frequency and $\beta_L =2\pi L I_c/\Phi_0$.
Subtracting and adding the voltage relations (\ref{voltagerel1eq}) and (\ref{voltagerel2eq}) in order to express them in terms of variables $\gamma_{\pm}$, we obtain
\begin{equation}
V^-_w -V^+_w=-\frac{M_m^2}{2L}\left(\dot{I}^-_w+\dot{I}^+_w\right)+\frac{M_m}{L}\frac{\Phi_0}{\pi}\frac{d}{dt}\left[\gamma_- +\pi\left( n+\frac{\Phi_{\mathrm{ext}}}{\Phi_0}\right)\right],
\label{voltdiffeq}
\end{equation}
where we have again used equations (\ref{phasefluxeq}) and (\ref{fluxeq}), and 
\begin{equation}
\dot{V}^-_w +\dot{V}^+_w=\frac{1}{2}\left(L-2M_m\right)\left(\ddot{I}^-_w -\ddot{I}^+_w\right)+\frac{2}{C_m}\left(I^-_w -I^+_w\right) +\frac{\Phi_0}{\pi}\ddot{\gamma}_+,
\label{voltsumeq}
\end{equation}
where we have taken time derivatives to replace the charges with currents and we have also used the current conservation relation (\ref{currentconseq}). Equations (\ref{gammapluseq}), (\ref{gammaminuseq}), (\ref{voltdiffeq}), (\ref{voltsumeq}) along with the wave equation (\ref{waveeq}) completely specify the classical dynamics of the coupled CPW-dc SQUID system.

We shall restrict ourselves to CPW mode solutions that reflect the symmetry of the circuit. In particular, we consider two types of solution: 1) Voltage antinode and current node at $z=0$: $V^-_w=V^+_w$, $I^-_w=-I^+_w$; 2) Voltage node and current antinode at $z=0$: $V^-_w=-V^+_w$, $I^-_w=I^+_w$. In the former case, the CPW-SQUID coupling is predominantly capacitive and results in the microwave resonator mode-charge qubit Hamiltonian, while in the latter case the coupling is inductive and results in the microwave resonator mode-flux qubit Hamiltonian. The following two subsections deal with each case in turn. 

\subsection{Voltage antinode equations: capacitive coupling}

For a voltage antinode-current node solution, we have from equation (\ref{voltdiffeq})  that $\gamma_-$ must be a constant. From equation (\ref{gammaminuseq}), we then require that $\gamma_- = q \pi$ and $\Phi_{\mathrm{ext}}=-(q+n)\Phi_0$, where $q$ is an arbitrary integer. Thus, we see that in order to have a non trivial solution for the remaining $\gamma_+$  variable, the external flux is constrained to be an integer multiple of the flux quantum. However, if $\beta_L\ll 1$, i.e., the SQUID loop self inductance $L$ is small, then to leading, zeroth order in $\beta_L$ a voltage antinode solution is allowed provided $\gamma_-=-\pi \left(n+\Phi_{\mathrm{ext}}/\Phi_0\right)$ and the external flux need not be constrained. Assuming therefore a small loop area SQUID with  $\beta_L\ll 1$ in the CPW resonator voltage antinode  case and neglecting the inductance dependent terms in equation (\ref{voltsumeq}), we obtain
\begin{equation}
\dot{V}^-_w=\frac{2}{C_m} I_w^- +\frac{\Phi_0}{2\pi}\ddot{\gamma}_+.
\label{voltsum2eq}
\end{equation}
Using equation  (\ref{voltsum2eq}), equation (\ref{gammapluseq}) for $\gamma_+$ thus becomes
\begin{equation}
\frac{\Phi_0}{2\pi}C_{\Sigma}\ddot{\gamma}_+ +2 I_c\cos\left(\pi\Phi_{\mathrm{ext}}/\Phi_0\right) \sin\gamma_+=C_m\dot{V}^-_w,
\label{gammaplus2eq}
\end{equation}
where $C_{\Sigma}=2 C_J+C_m$ and we have set $n=0$, since observable quantities do not depend on $n$. Equations (\ref{voltsum2eq}) and (\ref{gammaplus2eq}) along with the wave equation (\ref{waveeq}) completely specify the voltage antinode restricted classical dynamics.

The qubit control and readout protocols~\cite{armournjp08} require driving the CPW close to one of its resonant modes. In the voltage antinode case, this is usually  the second fundamental mode~\cite{blaispra04}. The subsequent analysis of the  CPW-SQUID dynamics is greatly simplified if we can first effectively replace the CPW with a single harmonic oscillator at the relevant microwave mode frequency. In \ref{singlemode}, we derive these approximate, effective single oscillator mode equations and also give the conditions under which these equations are valid.  In the voltage antinode second fundamental mode case, we obtain for the approximate coupled single oscillator-SQUID equations of motion:
\begin{equation}
C_{\Sigma}\frac{\Phi_0}{2\pi} \ddot{\gamma}_+ +2 I_c \cos(\pi\Phi_{\mathrm{ext}}/\Phi_0)\sin\gamma_+=C_m \frac{\Phi_0}{2\pi}\ddot{\phi}
\label{gammaplus3eq}
\end{equation}
and
\begin{equation}
\frac{C_w l}{2}\frac{\Phi_0}{2\pi}\left[\ddot{\phi}+\left(\frac{2\pi v}{l}\right)^2\phi\right]=C_m \frac{\Phi_0}{2\pi}\ddot{\gamma}_+,
\label{phiosceq}
\end{equation}
where $\phi(t)$ is the oscillator coordinate giving the CPW phase field amplitude at the SQUID location $z=0$ and where we have neglected an order $C_m/(C_w l)\ll 1$ shift in the CPW second fundamental mode frequency $\omega_w=2\pi v/l$.

Equations of motion (\ref{gammaplus3eq}) and (\ref{phiosceq}) follow from the Lagrangian
\begin{eqnarray}
L(\gamma_+,\phi,\dot{\gamma}_+,\dot{\phi})&=&\frac{1}{2}C_{\Sigma}\left(\frac{\Phi_0}{2\pi}\right)^2 \dot{\gamma}_+^2+\frac{I_c\Phi_0}{\pi}\cos(\pi\Phi_{\mathrm{ext}}/\Phi_0)\cos\gamma_+\cr
&&+\frac{1}{2}C_w l\left(\frac{\Phi_0}{2\pi}\right)^2\left[\frac{\dot{\phi}^2}{2}-\left(\frac{2\pi v}{l}\right)^2\frac{\phi^2}{2}\right]\cr
&&-C_m \frac{\Phi_0}{2\pi}\dot{\gamma}_+\left(\frac{\Phi_0}{2\pi}\dot{\phi}+V(t)\right),
\label{lagrangiancapeq}
\end{eqnarray}
where we have included a CPW voltage bias $V(t)$, assumed to be slowly varying in time as compared with the microwave mode frequency $\omega_w$. As we shall see in section~\ref{comparison}, this additional voltage bias is necessary in order to tune the qubit's operating point, couple strongly the qubit with the mechanical oscillator, and also drive the mechanical oscillator on resonance~\cite{armournjp08}.

Using the Lagrangian (\ref{lagrangiancapeq}) to construct the associated Hamiltonian, we have for the generalized momenta:
\begin{equation}
p_{\phi}=\frac{\partial L}{\partial \dot{\phi}}=\frac{C_wl}{2}\left(\frac{\Phi_0}{2\pi}\right)^2\dot{\phi}-C_m\left(\frac{\Phi_0}{2\pi}\right)^2\dot{\gamma}_+
\label{pphieq}
\end{equation}
and
\begin{equation}
p_{\gamma_+}=\frac{\partial L}{\partial \dot{\gamma}_+}=C_{\Sigma} \left(\frac{\Phi_0}{2\pi}\right)^2\dot{\gamma}_+-C_m\frac{\Phi_0}{2\pi}\left(\frac{\Phi_0}{2\pi}\dot{\phi}+V(t)\right).
\label{pgammaeq}
\end{equation}
Expressing the velocities in terms of the momenta and coordinate variables and substituting into the definition for the Hamiltonian, $H=p_{\phi}\dot{\phi} +p_{\gamma_+}\dot{\gamma}_+ -L$, we obtain after some algebra
\begin{eqnarray}
&&H(\gamma_+,\phi,p_{\gamma_+},p_{\phi})=C_{\Sigma}^{-1}\left(\frac{2\pi}{\Phi_0}\right)^2\frac{p_{\gamma_+}^2}{2}-\frac{I_c\Phi_0}{\pi}\cos(\pi\Phi_{\mathrm{ext}}/\Phi_0)\cos\gamma_+\cr
&&+\frac{2}{C_w l}\left(\frac{2\pi}{\Phi_0}\right)^2 \frac{p_{\phi}^2}{2}+\frac{1}{2} C_w l \left(\frac{\Phi_0}{2\pi}\right)^2 \left(\frac{2\pi v}{l}\right)^2 \frac{\phi^2}{2}\cr
&&+2\frac{C_m}{C_w l C_{\Sigma}}\left(\frac{2\pi}{\Phi_0}\right)^2 p_{\gamma_+} p_{\phi},
\label{hamiltoniacapeq}
\end{eqnarray}
where we have  have dropped overall constant (i.e., variable independent) terms.

From the definition (\ref{pgammaeq}) for the generalized momentum $p_{\gamma_+}$ and recalling that $\Phi_0 =h/(2 e)$, we see that
\begin{equation}
p_{\gamma_+}=-\hbar[N-N_m (t)],
\label{Neq}
\end{equation}
where $N$ is the number of excess Cooper pairs on the island represented by the directly connected nodes between the $C_m$ and $C_J$ capacitances (see figure~\ref{schemefig}) and $N_m (t) =-C_m V (t)/(2e)$ is the polarization charge induced by the slowly varying voltage bias on the $C_m$ electrode, expressed in units of Cooper pair number. Expression (\ref{Neq}) then suggests an alternative formulation of the Hamiltonian in terms of the Cooper pair number $N$:
\begin{eqnarray}
&&H(\gamma_+,\phi,N,p_{\phi})=\frac{(2e)^2}{2C_{\Sigma}} [N-N_m (t)]^2-\frac{I_c\Phi_0}{\pi}\cos(\pi\Phi_{\mathrm{ext}}/\Phi_0)\cos\gamma_+\cr
&&+\frac{2}{C_w l}\left(\frac{2\pi}{\Phi_0}\right)^2 \frac{p_{\phi}^2}{2}+\frac{1}{2} C_w l \left(\frac{\Phi_0}{2\pi}\right)^2 \left(\frac{2\pi v}{l}\right)^2\frac{\phi^2}{2}\cr
&&-4e\frac{C_m}{C_w l C_{\Sigma}}[N-N_m (t)]\frac{2\pi}{\Phi_0}p_{\phi}
\label{hamiltoniancap2eq}
\end{eqnarray}       
and we recognize in the top line of equation (\ref{hamiltoniancap2eq}) the usual Cooper pair box (CPB) Hamiltonian~\cite{makhlinrmp01}.   

\subsection{Voltage node equations: inductive coupling}

For a voltage node-current antinode solution, we see from equations (\ref{voltsumeq}) and (\ref{gammapluseq})  that $\gamma_+ =q \pi$, where $q$ is an arbitrary integer. Thus, equation (\ref{gammaminuseq}) for $\gamma_-$ becomes
\begin{equation}
\omega_J^{-2}\ddot{\gamma}_- +\sin\gamma_- +\frac{2}{\beta_L}\left(\gamma_-  +\frac{\pi\Phi_{\mathrm{ext}}}{\Phi_0}\right) =\frac{M_m}{L}\frac{I^-_w }{I_c} ,
\label{gammaminus2eq}
\end{equation}
where the mutual capacitance $C_m$ drops out and we  have set $n=q=0$. Equation (\ref{voltdiffeq}) also simplifies to
\begin{equation}
V^-_w=-\frac{M_m^2}{2L}\dot{I}^-_w+\frac{M_m}{L}\frac{\Phi_0}{2\pi}\dot{\gamma}_-,
\label{voltdiff2eq}
\end{equation}
where we have assumed that $\dot{\Phi}_{\mathrm{ext}}\approx 0$, i.e.,  changes slowly compared with the dynamical timescales of the SQUID and CPW. Equations (\ref{gammaminus2eq}) and (\ref{voltdiff2eq}) along with wave equation (\ref{waveeq}) completely specify the voltage node restricted classical dynamics.

The approximate,  first fundamental (i.e., lowest frequency) microwave mode-coupled SQUID equations that follow from equations (\ref{gammaminus2eq}) and (\ref{voltdiff2eq}) are (see~\ref{singlemode}):
\begin{equation}
C_J \frac{\Phi_0}{2\pi}\ddot{\gamma}_- +I_c \left[\sin \gamma_- +\frac{2}{\beta_L}\left(\gamma_-+\frac{\pi\Phi_{\mathrm{ext}}}{\Phi_0}\right)\right]=\frac{M_m}{L}  \frac{\Phi_0}{2\pi L_w}k_0\tan(k_0 l/2)\phi
\label{gammaminus3eq}
\end{equation} 
and
\begin{equation}
\frac{C_w l}{4} \frac{\Phi_0}{2\pi}\tan(k_0 l/2)\left[\ddot{\phi}+\left(\frac{\pi v}{l}\right)^2 \phi\right]=\frac{M_m}{L} \frac{\Phi_0}{2\pi L_w}k_0 \gamma_-,
\label{phiosc2eq}
\end{equation}
where from equation (\ref{phipeq}) in~\ref{singlemode} the term $\Phi_0/(2\pi L_w)k_0\tan(k_0 l/2)\phi$ appearing in the  $\gamma_-$ equation of motion (\ref{gammaminus3eq}) is just the CPW current at the SQUID location $z=0$. The wavenumber $k_0$ is the fundamental solution to the equation
\begin{equation}
\frac{k_0l}{2}\tan\left(\frac{k_0 l}{2}\right)=-\frac{L_w l L}{M^2_m}.
\label{trancendeq}
\end{equation}
Neglecting order $M_m/(L_w l)\ll1$ corrections, we have approximately for the fundamental wavenumber and mode frequency: $k_0= \pi/l$ and $\omega_w= \pi v/l$, respectively.

Equations of motion (\ref{gammaminus3eq}) and (\ref{phiosc2eq}) follow from the Lagrangian
\begin{eqnarray}
&&L(\gamma_-,\phi,\dot{\gamma}_-,\dot{\phi})=\cr
&&\frac{1}{2}C_J\left(\frac{\Phi_0}{2\pi}\right)^2\dot{\gamma}_-^2+\frac{I_c\Phi_0}{2\pi}\left[\cos\gamma_- -\beta_L^{-1}\left(\gamma_-+\frac{\pi}{\Phi_0}\left[\Phi_{\mathrm{ext}}-M_m I(t)\right]\right)^2\right]\cr
&&+\frac{1}{4}C_w l\left(\frac{\Phi_0}{2\pi}\right)^2\tan^2(k_0l/2)\left[\frac{\dot{\phi}^2}{2}-\left(\frac{\pi v}{l}\right)^2\frac{\phi^2}{2}\right]\cr
&&+\frac{M_m}{L}\frac{\Phi_0}{2\pi}\gamma_- \left[\frac{\Phi_0}{2\pi L_w}k_0\tan(k_0 l/2)\phi\right],
\label{lagrangianindeq}
\end{eqnarray}
where we have included a CPW current bias $I(t)$, assumed to be slowly varying in time as compared with the fundamental microwave mode frequency $\omega_w$. As we shall see in section~\ref{comparison}, this additional current bias is necessary in order couple strongly the qubit with the mechanical oscillator and also drive the mechanical oscillator on resonance~\cite{armournjp08}.

The Hamiltonian associated with Lagrangian (\ref{lagrangianindeq}) is
\begin{eqnarray}
&&H(\gamma_-,\phi,p_{\gamma_-},p_{\phi})=\cr
&&C_J^{-1} \left(\frac{2\pi}{\Phi_0}\right)^2 \frac{p_{\gamma_-}^2}{2}-\frac{I_c\Phi_0}{2\pi}\left[\cos\gamma_- -\beta_L^{-1}\left(\gamma_-+\frac{\pi}{\Phi_0}\left[\Phi_{\mathrm{ext}}-M_m I(t)\right]\right)^2\right]\cr
&&+\frac{4}{C_w l}\tan^{-2}(k_0l/2)\left(\frac{2\pi}{\Phi_0}\right)^2\frac{p_{\phi}^2}{2}
+\frac{1}{4}C_w l \left(\frac{\Phi_0}{2\pi}\right)^2\tan^2(k_0l/2) \left(\frac{\pi v}{l}\right)^2\frac{\phi^2}{2}\cr
&&-\frac{M_m}{L}\frac{\Phi_0}{2\pi}\gamma_- \left[\frac{\Phi_0}{2\pi L_w}k_0\tan(k_0 l/2)\phi\right].
\label{hamiltonianindeq}
\end{eqnarray}

\section{Quantum equations of motion}
\label{quantumequations}
The goal of this section is to derive the quantum Langevin equations for the coupled CPW microwave oscillator-charge(flux) qubit-mechanical oscillator system, taking into account the effects of damping and noise on the three subsystems. We shall begin by deriving the closed system quantum Hamiltonians, treating the (voltage antinode) capacitively-coupled charge qubit  and (voltage node) inductively-coupled flux qubit systems separately as we have done in the previous section.

\subsection{Charge qubit}
It is most convenient to work in terms of the CPW microwave oscillator raising and lowering operators:
\begin{equation}
\hat{a}_w^{\pm}=\frac{1}{\sqrt{2 m_w \omega_w\hbar}}\left(m_w \omega_w\hat{\phi}\mp i \hat{p}_{\phi}\right),
\label{apmeq}
\end{equation}
where, from equation (\ref{hamiltoniancap2eq}), $m_w=C_w l/2 (\Phi_0/2\pi)^2$ and $\omega_w=2\pi v/l$. 
The Hamiltonian (\ref{hamiltoniancap2eq}) then becomes
\begin{eqnarray}
H&=&\frac{(2e)^2}{2 C_{\Sigma}} [N-N_m(t)]^2-\frac{I_c\Phi_0}{\pi}\cos(\pi\Phi_{\mathrm{ext}}/\Phi_0)\cos\gamma_+ +\hbar\omega_w a_w^+ a_w\cr
&&-\frac{C_m}{C_{\Sigma}}\sqrt{ \hbar\omega_w\frac{(2e)^2}{C_w l}} [N-N_m(t)] i(a_w^+ -a_w),
\label{hamiltoniancap3eq}
\end{eqnarray}       
where for notational convenience we have dropped the hats and also the `-' superscript on the lowering operator. Examining the interaction Hamiltonian, we see that the coupling constant in units of $\hbar \omega_w$ depends essentially on the ratio of the CPW  single Cooper pair charging energy to the CPW mode  single photon energy~\cite{blaispra04}.

Note from equation (\ref{Neq}) that the standard Poisson bracket relation $\{\gamma_+,P_{\gamma_+}\}=1$ gives $\{\gamma_+,N\}=-\hbar^{-1}$ for the variables $\gamma_+$ and $N$. Thus, when we quantize using the correspondence principle between Poisson brackets and commutators, i.e., $\{\cdot,\cdot\}=c\rightarrow -i\hbar^{-1}[\cdot,\cdot]=c$ (for some constant $c$), we therefore have that $[\hat{\gamma}_+,\hat{N}]=-i$.

We now truncate the SQUID Hilbert space down to the lowest energy, two-dimensional Hilbert space, obtaining the charge qubit Hamiltonian. Working in the representation for which the number operator $\hat{N}$ is diagonal, the commutation relation $[\hat{\gamma}_+,\hat{N}]=-i$ gives $\hat{\gamma}_+ \leftrightarrow -i d/dN$. For the number operator, we have $\hat{N}=N|N\rangle\langle N| +(N+1) |N+1\rangle\langle N+1|=(N+\frac{1}{2})(|N\rangle\langle N|+|N+1\rangle\langle N+1|)-\frac{1}{2} (|N\rangle\langle N|-|N+1\rangle\langle N+1|)=(N+\frac{1}{2}) I -\frac{1}{2}\sigma_z$, where we have truncated to the two-dimensional Hilbert space spanned by eigenkets $|N\rangle\equiv \left({1}\atop{0}\right)$, $|N+1\rangle\equiv \left({0}\atop{1}\right)$ of the number operator; in order to be within this subspace, we require $V(t)$ to be tuned such that $N\leq N_m (t)\leq N+1$, where recall $N_m (t) =-C_m V (t)/(2e)$.  
 The truncated charging energy term can similarly be expressed as $[\hat{N}-N_m(t)]^2=(N-N_m)^2 |N\rangle \langle N| +(N+1-N_m)^2 |N+1\rangle\langle N+1| ) =[\delta N(t)]^2 +\frac{1}{4} +\delta N(t) \sigma_z$, where $\delta N(t)=N_m(t) -(N+\frac{1}{2})$.  
Writing $\cos(\hat{\gamma}_+)=(e^{i\hat{\gamma}_+}+e^{-i\hat{\gamma}})/2=(e^{d/dN}+e^{-d/dN})/2$, we have $\cos(\hat{\gamma}_+)=(|N\rangle\langle N+1|+|N+1\rangle\langle N|)/2=\sigma_x /2$. Substituting the above truncated approximations into equation (\ref{hamiltoniancap3eq}), we obtain for the truncated Hamiltonian:
\begin{eqnarray}
H&=&E_{C_{\Sigma}} \delta N(t) \sigma_z -\frac{1}{2}E_J(\Phi_{\mathrm{ext}})\sigma_x+\hbar\omega_w a_w^+ a_w+\hbar g_C (a_w +a^+_w)\sigma_z\cr
&&+2\hbar g_C (a_w +a^+_w)\delta N(t)+E_{C_{\Sigma}} [\delta N(t)]^2,
\label{fulltruncatedchargehamiltonianeq}
\end{eqnarray}
where
\begin{equation}
g_C=\omega_w\frac{C_m}{2 C_{\Sigma}}\sqrt{\frac{(2e)^2/(C_w l)}{\hbar\omega_w}}
\label{gccoupling}
\end{equation}
is the microwave mode-qubit coupling~\cite{blaispra04}, and    
where $E_{C_{\Sigma}}=(2e)^2/(2C_{\Sigma})$ is the single Cooper pair charging energy for the total capacitance $C_{\Sigma}=2C_J +C_m$ and   $E_J(\Phi_{\mathrm{ext}})=2 E_J^0 \cos (\pi\Phi_{\mathrm{ext}}/{\Phi_0})$ is the effective flux-dependent Josephson energy, with $E_J^0 = I_c\Phi_0/(2\pi)$ the Josephson energy of a single JJ. Note that we have also made the canonical replacement  $a_w\rightarrow i a_w$, $a^+_w\rightarrow -i a^+_w$.   

We now incorporate the mechanical degree of freedom, supposing that a segment of the SQUID loop directly opposite the centre conductor forms a mechanically compliant, doubly-clamped beam. For a small in plane displacement  with centre-of-mass amplitude $x$, the position dependent mutual capacitance is $C_m(d+x)\approx C_m(d) (1 -\eta_C x/d)$, where $d$ is the static equilibrium gap between the centre conductor and opposite facing SQUID loop segment and $\eta_C$ is a geometrical factor of order one accounting for the non-uniform beam displacement. Substituting this $C_m(d+x)$ into the Hamiltonian (\ref{fulltruncatedchargehamiltonianeq}) and assuming that $C_m\ll C_{\Sigma}$, we see that to leading order the coupling to the mechanical resonator arises through a modulation of the $V(t)$ induced polarization $N_m(t)$, as well as through a modulation of the microwave mode-qubit coupling $g_C$. Express the applied voltage bias as $V(t)=V_{dc} +V_{ac}(t)$, where $V_{dc}$ is a constant dc voltage and  $V_{ac}(t)$ is slowly varying as compared with the microwave mode frequency $\omega_w$.  We obtain for the mechanical resonator contribution to the Hamiltonian:
\begin{eqnarray}
H_m&=&\hbar\omega_m a_m^+ a_m+\lambda_C (a_m +a_m^+)\sigma_z -\hbar\tilde{x}_C g_C (a_w+a_w^+)(a_m+a_m^+)\sigma_z\cr
&&-\lambda_C \frac{C_m V_{ac}(t)}{e} (a_m+a_m^+),
\label{mechanicalhamiltoniancapeq}
\end{eqnarray}   
where $\tilde{x}_C=\eta_C x_{zp}/d$ with $x_{zp}$ the zero-point uncertainty and we assume that it is  the frequency $\omega_m$ fundamental flexural mode of the mechanical resonator that  predominantly couples to the qubit  with strength
\begin{equation}
 \lambda_C=\eta_C \frac{x_{zp}}{d}\frac{C_m}{C_{\Sigma}} {e V_{dc}}.
 \label{cpboscillatorcoupling}
 \end{equation}
Hamiltonian~(\ref{mechanicalhamiltoniancapeq}) neglects direct microwave-mechanical oscillator interaction terms, assumed to have a small effect on the coupled dynamics since the two modes are significantly off-resonance: $\omega_m\ll\omega_w$. We also assume that $|V_{dc}| \gg e/C_m$, necessary in order to achieve strong mechanical oscillator-qubit coupling, and that the mechanical oscillator driving voltage $|V_{ac}(t)|\ll e/C_m$. Later, we shall see that the typical large mechanical quality factors ensure that such ac driving  voltages are adequate for exciting the mechanical oscillator to sufficiently large amplitude as required by the qubit control protocol~\cite{armournjp08}. 
 
The control protocol also requires operating at the charge degeneracy point $\delta N=-C_m V_{dc}/(2e) -N =0$, where the CPB qubit coherence time is a maximum~\cite{ithierprb05}.  Thus, the full microwave-qubit-mechanical oscillator Hamiltonian becomes
\begin{eqnarray}
H&=&\frac{1}{2}\hbar\omega_a \sigma_z +\hbar\omega_w a^+_w a_w +\hbar\omega_m a^+_m a_m +\hbar g_C (a_w+a_w^+)\sigma_x\cr
&&+\lambda_C (a_m+a_m^+)\sigma_x -\hbar\tilde{x}_C g_C (a_w +a_w^+) (a_m+a_m^+)\sigma_x\cr
&&-\lambda_C \frac{C_m V_{ac}(t)}{e} (a_m+a_m^+),
\label{concisechargehamiltonianeq}
\end{eqnarray}
where we have changed the Pauli matrix basis: $\sigma_x\rightarrow-\sigma_z$, $\sigma_z\rightarrow -\sigma_x$, and $\omega_a=E_J(\Phi_{\mathrm{ext}})/\hbar$ is the qubit transition frequency.  From (\ref{concisechargehamiltonianeq}), we see that the mechanical oscillator couples to the qubit and microwave oscillator through two interaction terms of the form $(a_m+a_m^+)\sigma_x$ and  $(a_w +a_w^+) (a_m+a_m^+)\sigma_x$, respectively. Given the wide separation of timescales, $\omega_m\ll \omega_a$, an adiabatic approximation can be performed to show that the $(a_m+a_m^+)\sigma_x$ interaction term gives rise to  an approximate, dispersive interaction $a_m^+ a_m\sigma_z $ between the mechanical oscillator and qubit~\cite{buksprb06}. Thus, a dominant effect of the qubit on the mechanical oscillator is to shift its frequency up or down depending on the state of the qubit. By preparing the qubit in a superposition state, the mechanical oscillator is in turn driven into a superposition state via the dispersive interaction, as analyzed in reference~\cite{armournjp08}. Provided the frequency difference $|\omega_w -\omega_a |$ is much larger than the coupling $g_C$, the microwave-qubit interaction $(a_w +a_w^+)\sigma_x$  can similarly be replaced by an approximate dispersive interaction $a_w^+ a_w \sigma_z$~\cite{blaispra04}. Supposing that the qubit remains in its ground state, the microwave mode then exerts a ponderomotive-type force on the mechanical oscillator described by the interaction $a_w^+ a_w (a_m+a_m^+)$. Thus, passive cooling or alternatively amplification of the mechanical motion can in principle be implemented with appropriate red or blue detunings of the microwave drive frequency with respect to the microwave mode  frequency $\omega_w$~\cite{blencoweprb07}. We discuss these possibilities in section~\ref{comparison}.

\subsection{Flux qubit}

In this section, we obtain the truncated flux qubit Hamiltonian. While the analysis is well known and involves basic Schr\"{o}dinger wave mechanics, it does serve a  purpose in bringing to the fore the differences in the flux and charge qubit parameter regimes. Readers not interested in the  details of the derivation may skip directly to the resulting qubit Hamiltonian~(\ref{concisefluxhamiltonianeq}).

From equation (\ref{hamiltonianindeq}), the microwave mode  effective mass is 
\begin{equation}
m_w=\frac{1}{4}C_w l \left(\frac{\Phi_0}{2\pi}\right)^2\tan^2(k_0l/2)
\label{mweq}
\end{equation}
and the microwave mode fundamental frequency is $\omega_m =\pi v/l$.
In terms of the raising and lowering operators, the Hamiltonian (\ref{hamiltonianindeq}) is then
\begin{eqnarray}
H&=&C_J^{-1} \left(\frac{2\pi}{\Phi_0}\right)^2 \frac{p_{\gamma_-}^2}{2}-\frac{I_c\Phi_0}{2\pi}\left[\cos\gamma_- -\beta_L^{-1}\left(\gamma_-+\frac{\pi}{\Phi_0}\left[\Phi_{\mathrm{ext}}-M_m I(t)\right]\right)^2\right]\cr
&&+\hbar\omega_w a^+_w a_w  -\frac{M_m}{L} \sqrt{2 \hbar\omega_w \frac{(\Phi_0/2\pi)^2}{L_w l}} \gamma_- (a_w^+ +a_w).
\label{hamiltonianind2eq}
\end{eqnarray}
Examining the interaction Hamiltonian, we see that the coupling constant in units of $\hbar \omega_w$ depends essentially on the ratio of the CPW  single flux quantum addition energy to the CPW mode  single photon energy.

\begin{figure}[htbp]
\centering
\epsfig{file=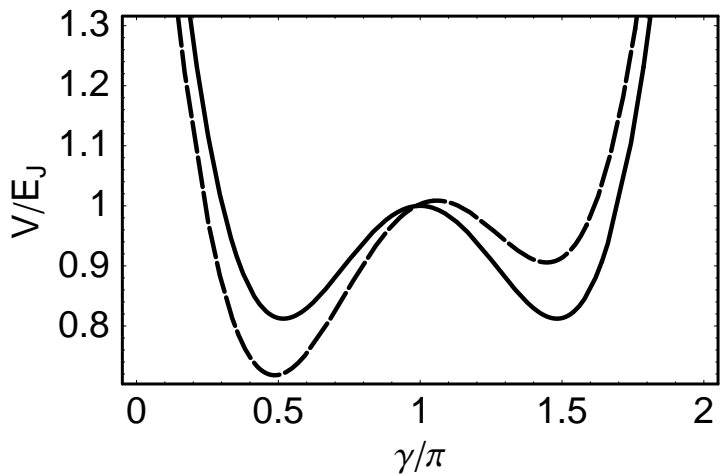,width=4.in} 
\caption{Double well potential for $\beta_L=3$ and $\Phi_{\mathrm{ext}}-M_m I_{dc}=-\Phi_0$ (solid line), $\Phi_{\mathrm{ext}}-M_m I_{dc}=-0.97 \Phi_0$ (dashed line).}  
\label{doublewell}
\end{figure}

The eigenstates and eigenvalues of the  SQUID part of the Hamiltonian~(\ref{hamiltonianind2eq}) can be obtained, e.g., by working in the Schr\"{o}dinger wavefunction form. In the representation $\hat{\gamma}\leftrightarrow \gamma$ and $\hat{p}_{\gamma}\leftrightarrow -i\hbar d/d\gamma$ (where we have dropped the `$-$' subscript on $\gamma$ for notational convenience), the SQUID Schr\"{o}dinger equation becomes 
\begin{eqnarray}
&&\hat{H}\Psi(\gamma)=\cr
&&-E_{C_J} \frac{d^2 \Psi(\gamma)}{d\gamma^2} +E_J \left[\beta_L^{-1} \left(\gamma +\frac{\pi}{\Phi_0}\left[\Phi_{\mathrm{ext}}-M_m I_{dc}\right]\right)^2 -\cos\gamma\right]\Psi(\gamma)\cr
&& =E\Psi(\gamma),
\label{schreq}
\end{eqnarray} 
where $E_{C_J}=(2e)^2/(2C_J)$ is the  Cooper pair charging energy  of the JJ capacitor,  $E_J=I_c\Phi_0/(2 e)$ is the Josephson energy, and we have suppressed for the present the slowly time-varying contribution $I_{ac}(t)$ to the current bias $I(t)=I_{dc}+I_{ac}(t)$. Provided $\beta_L=2\pi LI_c/\Phi_0\gtrsim 2$ and $\Phi_{\mathrm{ext}}-M_m I_{dc}=(2n+1)\Phi_0$, $n=0,\pm1, \pm2,...$, then the potential energy function admits a symmetric double well centered at $\gamma=-(2n+1)\pi$.  Figure~\ref{doublewell} shows an example plot  of the potential $V(\gamma)$ in region of its minimum, indicating a double-well. Also shown is the resulting asymmetric double well when the external flux $\Phi_{\mathrm{ext}}$ is detuned slightly from the above symmetric well condition value.  
In order that the lowest energy eigenvalues lie `within' the double well and hence be strongly anharmonic, we require that the harmonic ground state energy $E_0$  neglecting the cosine term in the potential $V(\gamma)$ be  smaller than the central barrier maximum ($=E_J$ for the degenerate, symmetric well---see figure~\ref{doublewell}). Thus, we require that $E_0=\sqrt{E_{C_J} E_J/\beta_L}<E_J$, or $E_{C_J}/E_J\ll\beta_L$. However, $E_{C_J}$ cannot be too small, for otherwise the energy spacing between the ground and first excited states will correspondingly be too small, making the resulting qubit susceptible to damping and decoherence by thermal and other low energy noise sources.

We have seen that, in contrast to the charge qubit (see section 2.1), the flux qubit requires a nonnegligible $\beta_L$, i.e, the flux qubit requires a much larger SQUID loop self-inductance $L$ than the CPB. Adequate effective self-inductances can be achieved by using small loop, multiple Josephson Junction SQUID devices~\cite{chiorescunature04}, where the additional JJ's compensate for the small loop area, hence avoiding potential increases in flux noise that come with using larger loop areas. However, in the present work, we consider only a double JJ, dc SQUID for simplicity; the relevant multiple JJ SQUID devices do not involve any conceptually new features and can be analyzed along similar lines. 

It is quite informative to solve the Schr\"odinger equation~(\ref{schreq}) approximately  in terms of the two harmonic ground basis functions for the quadratic well expansions about the two minima. We have:
\begin{equation}
V_i (\gamma)=V(\gamma_i)+E_J [\beta_L^{-1}+\cos(\gamma_i)/2] (\gamma-\gamma_i)^2,
\label{harmonicpoteq}
\end{equation} 
where the double well minima, $\gamma_i,~ i=1,2$, are solutions to
\begin{equation}
2\beta_L^{-1} \left(\gamma_i +\frac{\pi}{\Phi_0}\left[\Phi_{\mathrm{ext}}-M_m I_{dc}\right]\right) +\sin\gamma_i=0.
\label{wellmininimaeq}
\end{equation}
The normalized ground eigenfunctions of the harmonic approximation Hamiltonians with $V(\gamma)$ replaced by $V(\gamma_i)$ are
\begin{eqnarray}
\Psi_i(\gamma)&=&\left[\frac{E_J}{\pi^2 E_{C_J}}\left(\beta_L^{-1}+\cos(\gamma_i)/2\right)\right]^{1/8}\cr
&&\times\exp\left[-\frac{1}{2}\sqrt{\frac{E_J}{E_{C_J}}\left[\beta_L^{-1}+\cos(\gamma_i)/2\right]}(\gamma-\gamma_i)^2\right].
\label{harmonicfunctioneq}
\end{eqnarray}
The lowest two eigenfunctions of the double well Hamiltonian $\hat{H}$ are then approximately expressed as linear combinations of $|\Psi_i\rangle$: $|\Psi\rangle =A|\Psi_1\rangle +B|\Psi_2\rangle$, where the constants $A$, $B$ are solutions to
\begin{equation}
\left[
\begin{array}{cc}
  \langle\Psi_1|\hat{H}|\Psi_1\rangle-E&\langle\Psi_1|\hat{H}|\Psi_2\rangle-E\langle\Psi_1|\Psi_2\rangle  \\
 \langle\Psi_2|\hat{H}|\Psi_1\rangle-E\langle\Psi_2|\Psi_1\rangle   & \langle\Psi_2|\hat{H}|\Psi_2\rangle-E
\end{array}
\right] \left[
\begin{array}{c}
  A\\
 B
\end{array}\right]=
 \left[
\begin{array}{c}
 0\\
 0
\end{array}
\right]
\label{evaluefunctioneq}
\end{equation}
and the solutions $E$ give the approximate energy eigenvalues of $\hat{H}$. Note that the two basis functions are not orthogonal; the solutions are a reasonable approximation provided $|\langle\Psi_1|\Psi_2\rangle|\ll 1$.

Let us now consider a concrete example. Taking $L=500$~pH and $I_c=2.0~\mu{\mathrm{A}}$, we have $\beta_L=3.03$ and $E_J=4.1$~meV. Taking $C_J=1$~fF, we have $E_{C_J}=0.3$~meV, so that $E_{C_J}/(E_J\beta_L)=0.03\ll 1$. Considering a symmetric well  with $\Phi_{\mathrm{ext}}-M_m I_{dc}=(2n+1)\Phi_0$, we find that $A=\pm B$, and the approximate energy eigenvalues are: $E_-=3.95$~meV and $E_+=4.01$~meV with difference $\Delta E=E_+-E_-=55~\mu{\mathrm{eV}}$ (and $\Delta E/h=13~{\mathrm{GHz}}$). The approximate normalized eigenstates are $|\Psi_{\pm}\rangle=0.71 (|\Psi_1\rangle \pm |\Psi_2\rangle)$, i.e., (anti)symmetric combinations of the basis states, with the symmetric combination associated with the lower, ground eigenvalue $E_-$. For this example, we have $\langle\Psi_1|\Psi_2\rangle= 0.01$. On the other hand, for a slight external flux bias detuning $\Phi_{\mathrm{ext}}-M_m I_{dc}=(2n-0.98) \Phi_0$, making the double well potential asymmetrical, we have $E_- =3.75$~meV and $E_+=4.20$~meV with difference $\Delta E=E_+-E_-=450~\mu{\mathrm{eV}}$, and associated eigenstates $|\Psi_{-}\rangle=0.998|\Psi_1\rangle-0.058 |\Psi_2\rangle$ and $|\Psi_{+}\rangle=0.07|\Psi_1\rangle+0.998 |\Psi_2\rangle$. Thus, the lowest two eigenstates correspond to the phase being approximately localized to one or the other of the two wells.  

Proceeding formally with the truncation of the Hamiltonian~(\ref{hamiltonianind2eq})  to the subspace spanned by exact eigenstates $|\Psi_{\pm}\rangle$ and incorporating the mechanical degree of freedom through a position-dependent mutual inductance $M_m(d+x)\approx M_m(d) (1-\eta_L x/d)$ (with $\eta_L$ a geometrical factor of order one), we have:
\begin{eqnarray} 
&&H=\frac{1}{2}\Delta E \sigma_z +\hbar\omega_w a^+_w a_w +\hbar\omega_m a^+_m a_m -\hbar g_L (a_w +a_w^+)\gamma\cr
&&+\eta_L \frac{x_{zp}}{d}\frac{M_m}{L}\frac{\Phi_0}{2\pi} I_{dc} (a_m+a_m^+) \left(\gamma +\frac{\pi}{\Phi_0} [\Phi_{\mathrm{ext}}-M_m I_{dc}]\right) \cr
&&+\eta_L \frac{x_{zp}}{d}\frac{M_m}{L}\hbar g_L (a_w+a_w^+)(a_m+a_m^+) \gamma\cr
&&+\eta_L \frac{x_{zp}}{d}\frac{M_m}{L}\frac{\Phi_0}{2\pi} I_{ac}(t) (a_m+a_m^+)  \left(\gamma +\frac{\pi}{\Phi_0} [\Phi_{\mathrm{ext}}-2M_m I_{dc}]\right), 
\label{fulltruncatedfluxhamiltonianeq}
\end{eqnarray}
where the microwave mode-qubit coupling is
\begin{equation}
g_L= \omega_w\frac{M_m}{L}\sqrt{\frac{2(\Phi_0/2\pi)^2/(L_w l)}{\hbar\omega_w}}
\label{glcoupling}
\end{equation}
 and the truncated phase operator $\hat{\gamma}$  is
\begin{equation}
\hat{\gamma} = \gamma_{+-}\sigma_x +\frac{1}{2} (\gamma_{++}-\gamma_{--})\sigma_z+\frac{1}{2}(\gamma_{++}+\gamma_{--})I,
\label{gammatruncation}
\end{equation}
with $\gamma_{++}=\langle\Psi_+|\hat{\gamma}|\Psi_+\rangle$, etc.  Note that  $\Delta E$ and the truncated matrix elements $\gamma_{ij}$ are $\Phi_{\mathrm{ext}}$ and $I_{dc}$-dependent.    
For the example parameters considered above with symmetric double well potential [$\Phi_{\mathrm{ext}}-M_m I_{dc}=(2n+1)\Phi_0$], we have: $\gamma_{++}=\gamma_{--} =-(2n+1)\pi$, and $\gamma_{+-}\approx -\pi/2$, so that the $\sigma_z$ operator does not arise in the couplings. On the other hand, for the above asymmetric well example [$\Phi_{\mathrm{ext}}-M_m I_{dc}=(2n-0.98) \Phi_0$], we have $\gamma_{++}=4.6-2\pi n,  \gamma_{--} =1.6-2\pi n$, and $\gamma_{+-}=-0.2$, so that the $\sigma_z$ operator term is now present with much larger coupling as compared with the $\sigma_x$ operator term.      

We assume that the flux qubit's coherence time is optimized for the symmetric well case, i.e., the bias conditions are: $\Phi_{\mathrm{ext}}-M_m I_{dc}=(2n+1)\Phi_0$. The Hamiltonian~(\ref{fulltruncatedfluxhamiltonianeq}) then simplifies:
\begin{eqnarray}
&&H=\frac{1}{2}\hbar\omega_a \sigma_z +\hbar\omega_w a^+_w a_w +\hbar\omega_m a^+_m a_m - \hbar g_L \gamma_{+-}(a_w +a_w^+)\sigma_x\cr
&&+\lambda_L \gamma_{+-}(a_m+a_m^+)\sigma_x+\hbar\tilde{x}_L g_L\gamma_{+-} (a_w+a_w^+)(a_m+a_m^+)\sigma_x\cr
&&-\lambda_L \frac{M_m I_{ac}(t)}{(\Phi_0/\pi)} (a_m+a_m^+),
\label{concisefluxhamiltonianeq}
\end{eqnarray}
where $\omega_a=\Delta E/\hbar$, $\tilde{x}_L=\eta_L x_{zp}/d$, and the qubit-mechanical oscillator coupling is
\begin{equation}
\lambda_L=\eta_L \frac{x_{zp}}{d} \frac{M_m}{L} \frac{\Phi_0 I_{dc}}{2\pi}.
\label{fluxoscillatorcoupling}
\end{equation} 
In the Hamiltonian simplification~(\ref{concisefluxhamiltonianeq}), we neglect direct microwave oscillator-mechanical oscillator interaction terms and also assume small mechanical oscillator drive currents $|I_{ac}(t)|\ll\Phi_0/M_m$. 

Note that the microwave oscillator-flux qubit-mechanical oscillator Hamiltonian~(\ref{concisefluxhamiltonianeq}) is formally identical to the microwave oscillator-charge qubit-mechanical oscillator Hamiltonian~(\ref{concisechargehamiltonianeq}), the differences occuring only in their respective coupling strengths. This close identification will enable a straightforward comparison of the relative merits of the two Hamiltonians in section~\ref{comparison}.

\subsection{Open system quantum equations of motion}
We now include model environments for the CPW-qubit-mechanical resonator system. For the CPW, we suppose that the dominant direct dissipation mechanism is due to coupling to the microwave pump-probe line  used to drive and read out the state of the CPW.  
Assuming weak mechanical oscillator-bath and CPW mode-pump/probe line couplings, with the baths modeled as a continuous spectrum of free harmonic oscillators, we obtain the following Langevin equations for the mechanical and CPW mode operators $a_m$ and $a_w$, respectively:  
\begin{equation}
\frac{da_m}{dt}=-\frac{i}{\hbar} [a_m,H]-\gamma_m a_m -i\sqrt{2\gamma_m} e^{i\phi_b} a_b^{\mathrm{in}}
\label{ameq}
\end{equation} 
and
\begin{equation}
\frac{da_w}{dt}=-\frac{i}{\hbar} [a_w,H]-\gamma_w a_w -i\sqrt{2\gamma_w} e^{i\phi_p} a_p^{\mathrm{in}},
\label{aweq}
\end{equation}
where $H$ is the charge qubit Hamiltonian~(\ref{concisechargehamiltonianeq}) or flux qubit Hamiltonian~(\ref{concisefluxhamiltonianeq}). The $\gamma_i$'s are the various mode amplitude damping rates and the `in' and `out'  bath operators are related as follows \cite{gardinerpra85}:
\begin{equation}
a_i^{\mathrm{out}}(t)-a_i^{\mathrm{in}}(t)=-i\sqrt{2\gamma_i}e^{-i\phi_i} a_i (t).
\label{inoutrelation}
\end{equation}
This identity is used to determine the measured quantities' expectation values involving $a_p^{\mathrm{out}}(t)$ in  terms of the  solutions to the mode operators $a_i(t)$ and the prescribed initial $a_i^{\mathrm{in}}(t)$'s. 

The qubits themselves also  directly couple to environments other than the damped microwave and mechanical oscillators.  For completeness, Langevin equations describing the dissipative qubit dynamics are given in \ref{qubitlangevin}, although they are not used in the  present paper.

\section{Comparison of the flux and charge qubit implementations}
\label{comparison}
In the previous section, we derived  Hamiltonians describing a charge qubit~(\ref{concisechargehamiltonianeq}), or alternatively flux qubit~(\ref{concisefluxhamiltonianeq}), coupled to a microwave oscillator and mechanical oscillator. For convenience, we reproduce the two Hamiltonian expressions here:
\begin{eqnarray*}
H_{\mathrm{charge}}&=&\frac{1}{2}\hbar\omega_a \sigma_z +\hbar\omega_w a^+_w a_w +\hbar\omega_m a^+_m a_m +\hbar g_C (a_w+a_w^+)\sigma_x\cr
&&+\lambda_C (a_m+a_m^+)\sigma_x -\hbar\tilde{x}_C g_C (a_w +a_w^+) (a_m+a_m^+)\sigma_x\cr
&&-\lambda_C \frac{C_m V_{ac}(t)}{e} (a_m+a_m^+)
\end{eqnarray*}
and
\begin{eqnarray*}
&&H_{\mathrm{flux}}=\frac{1}{2}\hbar\omega_a \sigma_z +\hbar\omega_w a^+_w a_w +\hbar\omega_m a^+_m a_m - \hbar g_L \gamma_{+-}(a_w +a_w^+)\sigma_x\cr
&&+\lambda_L \gamma_{+-}(a_m+a_m^+)\sigma_x+\hbar\tilde{x}_L g_L\gamma_{+-} (a_w+a_w^+)(a_m+a_m^+)\sigma_x\cr
&&-\lambda_L \frac{M_m I_{ac}(t)}{(\Phi_0/\pi)} (a_m+a_m^+).
\end{eqnarray*} 
An important question concerns the relative merits of these two implementations. In this section, we address this  by comparing  realisable microwave-charge qubit and microwave-flux qubit coupling strengths $g_C$ (\ref{gccoupling}) and $g_L$ (\ref{glcoupling}),  qubit-mechanical oscillator coupling strengths $\lambda_C$ (\ref{cpboscillatorcoupling}) and $\lambda_L$ (\ref{fluxoscillatorcoupling}), and realisable mechanical displacement amplitudes due to achievable drive strengths $V_{ac}(t)$ and $I_{ac}(t)$.

Refering to the CPW circuit in figure~\ref{schemefig}, we consider a geometry modeled on the circuit QED device of reference~\cite{frunzioieee05} dictated by the requirement that the CPW impedance $Z=\sqrt{L_w/C_w}$ matches as closely as possible the typical $50\Omega$ impedances of coaxial feedlines. Furthermore, the resonant frequences of the first (voltage node at midpoint) and second (voltage antinode at midpoint) modes are required to be in the $5-10$~GHz range in order be in the quantum regime for the CPW at low dilution fridge temperatures, as well as to match the level separations of the flux or charge qubit.   We consider the centre conductor of the CPW to be uniformly $10~\mu$m wide and separated from the lateral ground planes by a $5~\mu$m gap. The doubly-clamped, suspended beam segment of the dc SQUID directly opposite the centre conductor is assumed to be $200$~nm wide, with a gap of $100$~nm between the two. All the described elements of the CPW and SQUID are $200$~nm thick. The substrate is assumed to be silicon, with an etched out trench beneath the mechanical resonator that extends up to the adjacent centre conductor edge.

The mutual capacitance per unit length between the centre conductor and lateral ground planes, $C_w$, is obtained as follows. Apply two linearly independent static voltage configurations to the centre and lateral planes and  for each configuration evaluate (numerically) the total stored electric field energy per unit length: $U=\frac{1}{2}\sum_{i,j=1}^2 V_i {C_w}_{i j} V_j$. Next, invert to obtain the capacitance per unit length matrix ${C_w}_{ij}$. Finally, obtain the mutual capacitance per unit length in terms of the capacitance matrix using
\begin{equation}
C_w=\frac{{C_w}_{11}{C_w}_{22}-{C_w}^2_{12}}{{C_w}_{11}+{C_w}_{22}+2{C_w}_{12}}.
\label{mutualcapacitanceeq}
\end{equation}
On the other hand, to obtain the self inductance per unit length of the centre inductor, $L_w$,  apply some non-zero steady state current  $I$ to the centre conductor, with zero applied current through the lateral conductors and then evaluate (numerically) the total stored magnetic field energy per unit length $U=\frac{1}{2} L_w I^2$. The self inductance is then simply $L_w=2 U/I^2$.  Carrying out these procedures, we obtain $C_w=2.01~{\mathrm{pF}/\mathrm{cm}}$ and $L_w=6.35~{\mathrm{nH}/\mathrm{cm}}$. These values give an impedance $Z=56\Omega$ and $v=(L_w C_w)^{-1/2}=8.85\times 10^7~{\mathrm{m}/\mathrm{s}}$ for the phase velocity. Therefore, the first mode frequency is $f_1 =v/(2l)=4.43~{\mathrm{GHz}/\mathrm{cm}}$ and second mode frequency is twice this: $f_2 =v/l=8.85~{\mathrm{GHz}/\mathrm{cm}}$. Thus, the CPW should be about $1~\mathrm{cm}$ long in order to have the  first two microwave modes in the desired frequency range. In order to evaluate the mutual capacitance $C_m$ and mutual inductance $M_m$ between the centre conductor and suspended SQUID beam, we assume for simplicity an infinitely long beam and  obtain: $C_m /l\approx 50~{\mathrm{aF}/\mu\mathrm{m}}$ and $M_m /l\approx 1~{\mathrm{pH}/\mu\mathrm{m} }$. 

For the microwave mode-qubit  coupling strengths, we also require the SQUID qubit JJ capacitance values (for the antinode case) and self inductance values (for the node case). As example values, we refer to the CPB qubits of the Yale group~\cite{wallraffnature04} and the flux qubits of the Delft group~\cite{chiorescunature04}. We shall consider only  order of magnitude estimates. Beginning with the inductively coupled flux qubit, we first note that the requirement $\beta_L\sim 1$ gives $L\sim\frac{\hbar}{2e I_c}$. Thus, in order that the SQUID  inductance $L$ and hence loop area not have to be too large, we require JJ's with large critical current $I_c$ values. The Delft group JJ's have $I_c\sim 1~\mu$A, giving $L\sim 100~{\mathrm{pH}}$. Considering a CPW length of $1~{\mathrm{cm}}$ gives a fundamental (node) mode frequency $v/(2l)\sim 5~{\mathrm{GHz}}$, comparable to the Rabi oscillation frequency at the symmetry point of the flux qubit in reference~\cite{chiorescunature04}. A few micron long mechanical resonator gives $M_m\sim 1~{\mathrm{pH}}$, so that we have for the inductance ratio $M_m/L\sim 0.01$.  For the capacitive coupling in the case of the CPB qubit,  a CPW length of $2~{\mathrm{cm}}$ gives a second (antinode) mode frequency $v/l\sim 5~{\mathrm{GHz}}$, comparable to the single electron charging energy (in units $h^{-1}$) of the original CPB qubit in reference~\cite{wallraffnature04}. Therefore, $C_{\Sigma}\sim 1~{\mathrm{fF}}$ and a few micron long mechanical resonator gives a  coupling ratio $C_m/C_{\Sigma}\sim 0.1$. 

With all the capacitance and inductance values in hand, we are now ready to compare the coupling strengths $g_C$ and $g_L$. We have:
\begin{equation}
g_L=\omega_w \frac{M_m}{L} \sqrt{\frac{{2(\Phi_0/2\pi)^2}/{(L_w l)}}{\hbar\omega}}\sim 0.01\omega_w
\label{indcoupstrengtheq}
\end{equation}
and
\begin{equation}
g_C=\omega_w \frac{C_m}{C_{\Sigma}}\sqrt{\frac{(2e)^2/(2C_w l)}{\hbar\omega}}\sim 0.01\omega_w,
\label{capcoupstrengtheq}
\end{equation}
where note that the CPW is twice as long ($2l$) as in the voltage node case ($l$) so that the 2nd mode and 1st mode frequencies are the same, allowing a direct comparison. 
Interestingly, although the inductance ratio $M_m/L$ is about an order of magnitude smaller than the capacitance ratio $C_m/C_{\Sigma}$, the flux quantum addition energy term is about an order of magnitude larger than the charging energy term, bringing the two coupling strength terms (\ref{indcoupstrengtheq}) and (\ref{capcoupstrengtheq}) into line. The reason for the difference in the flux quantum addition energy and charging energy magnitudes can be seen more clearly by taking the ratio of the square root terms:
\begin{equation}
\sqrt{\frac{{2(\Phi_0/2\pi)^2}/{(L_w l)}}{(2e)^2/(2C_w l)}}=\frac{h/e^2}{4\pi \sqrt{L_w/C_w}}=\frac{R_K}{4\pi Z}\approx 37.
\label{ratioeq}
\end{equation}   
Note that this is essentially just the ratio of the universal quantum of resistance (i.e., von Klitzing constant  $R_K=25.8~{\mathrm{k}}\Omega$)  to the CPW impedance and so with typical microwave device impedances in the several tens of Ohms together with the $4\pi$ factor in the denominator, we expect this ratio to be typically much larger than one. Thus,  with regards to the microwave mode-qubit coupling strengths, there is no advantage to be gained by implementing one qubit scheme over the other. 

Comparing the  qubit-mechanical oscillator coupling strengths (\ref{cpboscillatorcoupling}) and (\ref{fluxoscillatorcoupling}), we have
\begin{equation}
\frac{\lambda_C}{\lambda_L}=4\pi\frac{\eta_C}{\eta_L} \frac{(C_m/C_{\Sigma})}{(M_m/L)} \frac{ (V_{dc}/I_{dc})}{R_K}\sim 5 \times 10^{-3} \frac{(V_{dc}/I_{dc})}{1\ \Omega},\label{mechcouprat}
\end{equation}
where we have assumed  geometrical factors $\eta_C, \eta_L \sim 1$. Thus, applying a CPW current bias $I_{dc}\sim 5~{\mathrm{mA}}$ in the flux qubit device gives the same order mechanical coupling strength as applying a CPW voltage bias $V_{dc}\sim 1~{\mathrm{V}}$ in the charge qubit device. For a 5 micron long mechanical resonator with fundamental flexural frequency $\omega_m/(2\pi)=50~{\mathrm{MHz}}$, zeropoint displacement $x_{zp}=10^{-14}~{\mathrm{m}}$, and mechanical beam-centre conductor gap $d=100~{\mathrm{nm}}$, we have for the absolute, charge qubit-mechanical coupling strength
\[
\kappa_C=\frac{\lambda_C}{\hbar\omega_m}=\eta_C \frac{x_{zp}}{d}\frac{C_m}{C_{\Sigma}} \frac{e V_{dc}}{\hbar\omega_m}\sim 0.1 \frac{V_{dc}}{1{\mathrm{ V}}}.
\]
Therefore, a voltage bias $V_{dc}\sim 1~{\mathrm{V}}$  provides sufficiently strong charge qubit-mechanical oscillator coupling strength $\kappa_C$ for the scheme of reference~\cite{armournjp08}.

The $V_{ac}(t)$ and $I_{ac}(t)$ terms in Hamiltonians~(\ref{concisechargehamiltonianeq}) and (\ref{concisefluxhamiltonianeq}) are used to drive the mechanical oscillator into a non-zero displacement amplitude state, as required by the scheme of the companion paper~\cite{armournjp08} to rapidly generate a mechanical superposition state. Working with Hamiltonian~(\ref{concisechargehamiltonianeq}) and assuming a sinusoidal voltage drive that is resonant with the mechanical frequency, the steady state mechanical amplitude $x_m^0$ in units of the zeropoint displacement is
\begin{equation}
|\alpha_0|=\frac{x_m^0}{x_{zp}}=\eta_C \frac{x_{zp}}{d}\frac{C_m}{C_{\Sigma}}\frac{C_m |V_{dc}| V_{ac}^0}{\hbar \omega_m}Q_m,
\label{chargeamplitude}
\end{equation}
where $V_{ac}^0$ is the ac voltage drive amplitude and $Q_m$ is the mechanical quality factor. For the same mechanical oscillator parameters as above and quality factor $Q_m=10^4$ and dc voltage bias $V_{dc}=1~{\mathrm{V}}$, we have
\begin{equation}
|\alpha_0|\sim 10^7 \frac{V_{ac}^0}{1{\mathrm{ V}}}.
\label{alpha0charge}
\end{equation}
Thus, according to the mechanical superposition generation scheme~\cite{armournjp08}, ac voltage drive amplitudes of no more than a few microvolts are sufficient to obtain the required displacement amplitudes. Such ac voltage amplitudes do not significantly perturb the dc bias conditions on the CPB, since they  are considerably smaller than the voltage bias difference $2e/C_m\approx 0.5~{\mathrm{mV}}$ separating neighbouring CPB charge degeneracy points. 

Alternatively, working with the Hamiltonian~(\ref{concisefluxhamiltonianeq}) and assuming a sinusoidal current drive that is resonant with the mechanical frequency, the steady state mechanical amplitude $x_m^0$ in units of the zeropoint displacement is
\begin{equation}
|\alpha_0|=\frac{x_m^0}{x_{zp}}=\eta_L \frac{x_{zp}}{d}\frac{M_m}{L}\frac{M_m |I_{dc}| I_{ac}^0}{\hbar \omega_m}Q_m,
\label{fluxamplitude}
\end{equation}
where $I_{ac}^0$ is the ac current drive amplitude. Using the same mechanical resonator parameters as above and a dc current bias $I_{dc}=5~{\mathrm{mA}}$ (for which the flux qubit-mechanical oscillator coupling strength $\lambda_L$ coincides with  charge qubit-oscillator strength $\lambda_C$ -- see above), we have
\begin{equation}
|\alpha_0|\sim 10^6 \frac{I_{ac}^0}{1{\mathrm{  A}}}.
\label{alpha0flux}
\end{equation}
Thus, ac current drive amplitudes of no more than a few tens of microamps are sufficient to obtain the required displacement amplitudes for the mechanical superposition generation scheme~\cite{armournjp08}.  Such ac current amplitudes do not significantly perturb the dc bias conditions on the flux qubit, since they  are considerably smaller than the current bias difference $2\Phi_0/M_m\approx 1~{\mathrm{mA}}$ separating neighbouring symmetric double well configurations.

The $\hbar\tilde{x} g (a_w+a_w^+)(a_m+a_m^+)\sigma_x$ interaction terms in Hamiltonians~(\ref{concisechargehamiltonianeq}) and (\ref{concisefluxhamiltonianeq}) can in principle be employed to passively cool the mechanical oscillator (i.e., flexural mode) by driving the microwave mode off-resonance on the red-detuned side~\cite{blencoweprb07}. Provided $g/|\omega_a-\omega_w|\ll1$, the described interaction term can be approximately replaced with the dispersive interaction:
\begin{equation}
 \hbar\tilde{x} g (a_w+a_w^+)(a_m+a_m^+)\sigma_x\rightarrow\frac{\hbar\tilde{x} g^2}{|\omega_a-\omega_w|}a_w^+a_w (a_m+a_m^+)\sigma_z.
 \label{dispersive}
 \end{equation}
 If the qubit is prepared and subsequently remains in its ground state during the driving interval, then we obtain the familiar ponderomotive interaction between the microwave and mechanical oscillator modes.
 For the above considered parameter values, we have for the dispersive coupling strength in either qubit case: 
 \begin{equation}
 \frac{\tilde{x} g^2}{|\omega_a -\omega_w| \omega_w}\ll \frac{\tilde{x} g}{\omega_w}\sim 10^{-9}.
 \label{dispersivecouplingest}
 \end{equation}
 Interestingly, this upper bound on the achievable coupling strength is still orders of magnitude smaller than what is possible in a related scheme where the SQUID with mechanically compliant loop segment is instead imbedded within the planar waveguide, interrupting the centre stripline conductor~\cite{blencoweprb07}. Applying the theoretical analysis of reference~\cite{blencoweprb07} assuming a dispersive coupling strength $\sim 10^{-10}$, microwave mode of $5~{\mathrm{GHz}}$ and quality factor $Q_w=10^4$, and  mechanical flexural mode of $50~{\mathrm{MHz}}$ (thus we are in the good cavity limit: $\omega_m/\gamma_w\gg1$~\cite{marquardtprl07}) and quality factor $Q_m=10^4$, we find that  pump drive powers of magnitude $\sim 10~{\mathrm{mW}}$ are required in order to cool  the flexural mode down to a few energy quanta starting from a temperature of, say, $100~{\mathrm{mK}}$. Such drive power strengths are unrealistic given the need to substantially attenuate the thermal Johnson noise in the original GHz drive signal entering the RF line at room temperature~\cite{schusterphd07}. Thus, the dispersive coupling  in the presently considered scheme is not strong enough in order to achieve significant passive cooling. Nevertheless, as emphasized in the companion paper~\cite{armournjp08}, there is no upper thermal threshold to verifying mechanical superposition states, although for the considered device parameters and temperatures higher than a few tens of mK, we expect  it to become difficult to tease out the recoherences in an echo experiment that are the signature of mechanical superposition states. As long as the mechanical resonator temperature can be maintained to within a few tens of mK   (i.e., dilution fridge temperatures) during the drive and subsequent qubit control stages,  there should be no problem observing mechanical superposition states; cooling is  not essential for the scheme.           

\section{Conclusion} 

We have described an implementation of the  nanomechanical quantum  
superposition generation and verification scheme presented in  the  
companion reference~\cite{armournjp08}. The implementation is based on the  
circuit QED set-up with the incorporation of a mechanical degree of  
freedom formed out of a suspended SQUID loop segment located opposite  
the centre conductor of a coplanar waveguide. Two qubit  
realizations  were investigated, namely the capacitively coupled charge qubit  
with voltage-biased qubit-mechanical resonator coupling, and the inductively  
coupled flux qubit with current-biased qubit-mechanical resonator coupling. Both  
qubit realizations were found to have comparable and feasible resonator  
coupling strengths that are adequate for the quantum superposition  
generation scheme~\cite{armournjp08}. Ultimately, less  predictable qubit  
properties such as their  coherence times  will decide which  
implementation is the more promising of the two. Nevertheless, the present  
considerations suggest that both implementations are equally worth pursuing  
in experiment.

\ack
We especially thank A. Rimberg  and also  E. Buks, J. Hertzberg, and K. Schwab for  very helpful discussions. MPB was partly supported by the National Science Foundation (NSF) under NIRT Grant No. CMS-040431, the US-Israel Binational Science Foundation (BSF) and by the Foundational Questions Institute (FQXi). ADA was supported by the EPSRC under grant EP/E03442X/1.
\appendix
\section{Derivation of the single microwave mode-SQUID equations}
\label{singlemode}
The most general solution to the wave equation (\ref{waveeq}) for $\phi(z,t)$ that satisfies the boundary conditions $I_w(\pm l/2,t)=0$ takes the following Fourier integral form for $-l/2<z<0$:
\begin{eqnarray}
&&\phi^- (z,t)=\int dk A(k) \cos(kvt) \left[\cos(kz+\theta(k))+\cos(kz+kl-\theta(k))\right]\cr
&&+\int dkB(k)  \sin(kvt) \left[\sin(kz+\theta(k))-\sin(kz+kl-\theta(k))\right],
\label{phisolneq}
\end{eqnarray} 
with $\phi^+ (z,t)=\pm\phi^-(-z,t)$, $0<z<l/2$, for a voltage antinode (node) solution. Substituting Eq. (\ref{phisolneq}) evaluated at $z=0$ into Eqs. (\ref{voltsum2eq}), (\ref{gammaplus2eq}), (\ref{gammaminus2eq}), and (\ref{voltdiff2eq})  
can serve as a starting point for analyzing the classical dynamics of the coupled CPW-SQUID system. 
Because of the nonlinear ``force" terms in the SQUID part of the equations of motion (\ref{gammaplus2eq}) and (\ref{gammaminus2eq})  for $\gamma_\pm$, the coupled dynamics is expected to be quite nontrivial and rich in general. On the other hand, if the force terms had been linear, then a simpler normal mode analysis could be used with only two distinct $k$ values in (\ref{phisolneq}) necessary for constructing the normal mode solutions. Furthermore, provided the two normal mode $k$ values differ by much less than $l^{-1}$ [so that the dispersive differences in the $z$-dependent terms of (\ref{phisolneq}) can be neglected] , then the wave equation describing the CPW can be approximately replaced by a much simpler, single mode harmonic oscillator equation with linear coupling to the linearized SQUID. In other words, the CPW-linearized SQUID system can be modelled simply as two coupled oscillators. Nevertheless, under conditions of weak coupling between the CPW and SQUID, i.e.,  $C_m\ll C_{\Sigma}, C_w l$ and $M_m \ll L, L_w l$, it may still be possible to approximately replace the CPW with a single microwave frequency harmonic oscillator bilinearly coupled to the nonlinear SQUID. We shall assume this is the case and will now proceed in this appendix to derive the approximate  single harmonic oscillator-SQUID coupled equations of motion for the voltage antinode, capacitively coupled case.  We stress, however, that there still remains the issue of establishing the full domain of accuracy of the solutions to these approximate equations by comparing with the classical solutions assuming the full multimode Fourier integral form (\ref{phisolneq}).

Restricting to single mode forms of equation (\ref{phisolneq}), we have
\begin{equation}
\phi^-(z,t)=2A_0\cos[k(z+l/2)]\cos(kvt),
\label{phifieldeq}
\end{equation}
where we have fixed the phase in equation (\ref{phisolneq}) to be $\theta(k)=kl/2$. Defining $\phi_0=2A_0\cos(kl/2)$, the phase amplitude at the SQUID location $z=0$, we have
\begin{equation}
\phi^-(0,t)=\phi_0\cos(kvt)=\phi(t)
\label{phieq}
\end{equation}
and
\begin{equation}
\left.\frac{\partial\phi^-(z,t)}{\partial z}\right|_{z=0}=-k \tan(kl/2) \phi_0\cos(kvt)=-k \tan(kl/2) \phi(t),
\label{phipeq}
\end{equation}
Substituting equations (\ref{phieq}) and (\ref{phipeq}) into equation (\ref{voltsum2eq}) and  linearized equation (\ref{gammaplus2eq}) and expressing the latter two using more concise notation, we have
\begin{equation}
\ddot{\gamma}_+ +\alpha\gamma_+=\epsilon\ddot{\phi}
\label{gammaconciseeq}
\end{equation}
and
\begin{equation}
\frac{\epsilon}{2} \ddot{\phi}-\beta k\tan(kl/2)\phi=\frac{\epsilon}{2}\ddot{\gamma}_+,
\label{phiconciseeq}
\end{equation}
where $\ddot{\phi}=-\omega^2\phi=-(kv)^2 \phi$,  $\alpha=4\pi I_c\cos(\pi\Phi_{\mathrm{ext}}/\Phi_0)/(\Phi_0 C_{\Sigma})$, $\beta=(C_{\Sigma} L_w)^{-1}$ and $\epsilon=C_m/C_{\Sigma}\ll1$.
From the form of Eqs. (\ref{gammaconciseeq}) and (\ref{phiconciseeq}), we make a guess that the coupled oscillator equations of motion that yield approximately the same solutions take the form:
\begin{eqnarray}
\ddot{\gamma}_+ +a \gamma_+&=& b\epsilon \ddot{\phi}\cr
\ddot{\phi}+c\phi&=&d\epsilon\ddot{\gamma}_+.
\label{canonicalosceq}
\end{eqnarray}
However, because of the $\tan(kl/2)$ term in Eq. (\ref{phiconciseeq}), it will only be possible to match the solutions to those of the coupled oscillator equations with to-be-determined constant coefficients $a$, $b$, $c$ and $d$, provided that the two normal mode frequencies are sufficiently close to each other so that dispersive differences can be neglected.

In order to determine the oscillator equation coefficients in terms of $\alpha$, $\beta$ and $\epsilon$, we match the normal mode frequencies of Eqs. (\ref{gammaconciseeq}) and (\ref{phiconciseeq}) to those of Eq. (\ref{canonicalosceq}).  Solving first for the normal frequencies of Eq. (\ref{canonicalosceq}) to second order in $\epsilon$, we obtain 
\begin{equation}
\omega^+=k^+ v=\sqrt{a} \left(1+\frac{c}{2}\frac{bd}{a-c}\epsilon^2\right)
\label{omegapluseq}
\end{equation}
and
\begin{equation}
\omega^-=k^- v=\sqrt{c} \left(1-\frac{a}{2}\frac{bd}{a-c}\epsilon^2\right).
\label{omegaminuseq}
\end{equation}
The secular equation following from  Eqs. (\ref{gammaconciseeq}) and (\ref{phiconciseeq}) is
\begin{equation}
\left(\alpha-\omega^2\right) \left[\frac{\omega^2}{2}  \epsilon +\frac{\beta\omega}{v}\tan\left(\frac{\omega l}{2 v}\right)\right]+\frac{\omega^4}{2}\epsilon^2=0,
\label{seculareq}
\end{equation}
where we have used $k=\omega/v$. Solving Eq. (\ref{seculareq}) for the smallest mode $\omega$ to second order in $\epsilon$, we obtain:
\begin{equation}
\omega^+=\sqrt{\alpha}+\frac{\alpha v}{4\beta}\cot\left(\frac{\sqrt{\alpha}l}{2v}\right)\epsilon^2
\label{omegaplus2eq}
\end{equation}
and
\begin{equation}
\omega^- =\frac{2\pi v}{l}\left(1-\frac{v^2}{\beta l}\epsilon \right)-\frac{\left(2\pi v/l\right)^3 v^2/(\beta l)}{\alpha-(2\pi v/l)^2}\epsilon^2.
\label{omegaminus2eq}
\end{equation}
Comparing Eqs. (\ref{omegaplus2eq}) and (\ref{omegaminus2eq}) with Eqs. (\ref{omegapluseq}) and (\ref{omegaminuseq}), we can clearly see that the CPW-linearized SQUID mode frequency $\omega^+$ has a different form than the $\omega^+$ arising from the coupled oscillator equations (\ref{canonicalosceq}). This is because the boundary conditions on the CPW result in mode solutions with nonlinear dispersion, as signified by the presence of trigonometric terms. However, if  we assume that the modes $\omega^+$ and $\omega^-$ are sufficiently close to each other, then we can expand the cotangent term in the mode frequency difference. Eq. (\ref{omegaplus2eq}) then becomes approximately
\begin{equation}
\omega^+ =\sqrt{\alpha}+\frac{\left(2\pi v/l\right)^3 v^2/(\beta l)}{\alpha-(2\pi v/l)^2}\epsilon^2.
\label{omegaplus3eq}
\end{equation}
Validity of the second order in $\epsilon$ expansion together with this expansion in the frequency difference demands the following condition:
\begin{equation}
\frac{C_m^2}{C_{\Sigma}C_w l}l^{-1}\ll |k^+ -k^-| \ll l^{-1}.
\label{validitycondeq}
\end{equation}
The CPW-linearized SQUID mode frequencies now take the same form as those for the coupled harmonic oscillators. Matching to determine the oscillator equation coefficients, we have:
$a=\alpha$, $c=(2\pi v/l)^2 (1-2v^2\epsilon/(\beta l))$, $b=1$, and $d=2 v^2/(\beta l)$. Substituting these values into Eq. (\ref{canonicalosceq}) and restoring the nonlinear $\gamma_+$ force term, we finally obtain the desired coupled single oscillator-SQUID equations of motion~(\ref{gammaplus3eq}) and (\ref{phiosceq}). From Eq. (\ref{validitycondeq}), we assume these equations are a good approximation to the full equations provided the difference in the characteristic frequencies for the SQUID and CPW dynamics, $\delta\omega$, satisfies
\begin{equation}
 \frac{C_m^2}{C_{\Sigma}C_w l}\frac{v}{l}\ll |\delta\omega| \ll \frac{v}{l}.
\label{validitycond2eq}
\end{equation}

The derivation of the single oscillator-SQUID equations for the voltage node, inductively coupled case run along similar lines, yielding equations~(\ref{gammaminus3eq})  and (\ref{phiosc2eq}). We assume these equations are a good approximation to the full equations provided the difference in the characteristic frequencies for the SQUID and CPW dynamics, $\delta\omega$, satisfies
\begin{equation}
\frac{C_w l}{C_J}\left(\frac{M_m}{L}\right)^2 \frac{v}{l}\ll |\delta\omega| \ll \frac{v}{l}.
\label{validitycond3eq}
\end{equation}

\section{Derivation of the qubit Langevin equations}
\label{qubitlangevin}
In this appendix, we derive the qubit Langevin equations. This operator formulation provides a complementary approach for  addressing the dissipative qubit dynamics  to that of the more commonly employed Fokker-Planck, master equation approach which solves for the time-dependence of the qubit density matrix~\cite{armournjp08}. The Langevin equation formulation of the qubit dynamics naturally belongs to  the `in-out' quantum optics approach to quantum measurement and control~\cite{gardinerpra85} and can be straightforwardly applied to analyzing circuit QED set-ups~\cite{blencoweprb07}.
The qubit baths are modeled as a dense spectrum of oscillator modes and a rotating wave approximation is made for the assumed weak qubit-bath couplings:
\begin{eqnarray}
H_{\mathrm{qb-bath}}/\hbar&=& \int d\omega\omega a_1^+(\omega) a_1 (\omega) +\int d\omega\omega a_2^+(\omega) a_2 (\omega)\cr
&&+\int d\omega [K_1^*(\omega) a_1^+(\omega)\sigma^- +K_1(\omega)\sigma^+ a_1(\omega)]  \cr
&&+\int d\omega [K_2^*(\omega) a_2^+(\omega) \sigma_z +K_2(\omega) \sigma_z a_2(\omega)],
\label{qbbatheq}
\end{eqnarray}
where $\sigma^+$ and $\sigma^-$ are the qubit (spin) raising and lowering operators ($\sigma_x=\sigma^+ +\sigma^-$) and the $a_1$, $a_2$ are independent oscillator bath modes,   the first coupling $K_1$ induces qubit decay through spin flips and where  the second coupling $K_2$ causes pure dephasing. 
The equations for the bath modes are
\begin{eqnarray}
\frac{d a_1}{dt}&=&-i\omega a_1 -i K_1^* \sigma^-\cr
\frac{d a_2}{dt}&=&-i\omega a_1 -i K_2^* \sigma_z.
\label{batheqs}
\end{eqnarray}
For the qubit, it suffices to consider the equations for $\sigma_z$ and $\sigma^+$:
\begin{equation}
\frac{d\sigma_z}{dt}=-\frac{i}{\hbar}[\sigma_z,H]-2i\int d\omega (K_1 \sigma^+ a_1-K_1^* a_1^+  \sigma^-)
\label{sigmazeq}
\end{equation} 
and
\begin{equation}
\frac{d\sigma^+}{dt}=-\frac{i}{\hbar}[\sigma^+,H]-i\int d\omega K_1^* a_1^+ \sigma_z +2i\int d\omega (K_2^* a_2^+ \sigma^+ +K_2\sigma^+ a_2 ),
\label{sigmapeq}
\end{equation} 
where $H$ is the charge qubit Hamiltonian~(\ref{concisechargehamiltonianeq}) or flux qubit Hamiltonian~(\ref{concisefluxhamiltonianeq}). 
Integrating the bath equations of motion~(\ref{batheqs}), we have
\begin{eqnarray}
a_1(\omega,t)&=&-i K_1^*(\omega)\int_{t_0}^t dt' e^{-i\omega (t-t')}\sigma^- (t') +e^{-i\omega (t-t_0)} a_1 (\omega,t_0)\cr
a_2(\omega,t)&=&-i K_2^*(\omega)\int_{t_0}^t dt' e^{-i\omega (t-t')}\sigma_z (t') +e^{-i\omega (t-t_0)} a_2 (\omega,t_0).
\label{bathsoln}
\end{eqnarray}
Substituting these solutions into equations (\ref{sigmazeq}) and (\ref{sigmapeq}) for $\sigma_z$ and $\sigma^+$, we have
\begin{eqnarray}
\frac{d\sigma_z}{dt}&=&-\frac{i}{\hbar}[\sigma_z,H]\cr
&&-2\int_{t_0}^t dt' \left[\sigma^+(t)\sigma^-(t') \int d\omega |K_1|^2 e^{-i\omega (t-t')}\right.\cr
&&\left.+\sigma^+(t')\sigma^-(t) \int d\omega |K_1|^2 e^{+i\omega (t-t')}\right]\cr
&&-2i \sigma^+ (t)\int d\omega K_1 e^{-i\omega (t-t_0)} a_1(\omega,t_0)\cr
&&+2i \int d\omega K_1^* e^{+i\omega (t-t_0)} a^+_1(\omega,t_0) \sigma^- (t)
\label{zeffectivebatheq}
\end{eqnarray}
and
\begin{eqnarray}
\frac{d\sigma^+}{dt}&=&-\frac{i}{\hbar}[\sigma^+,H]+\int_{t_0}^t dt' \sigma^+(t') \sigma_z (t) \int d\omega |K_1|^2 e^{i\omega (t-t')}\cr
&& -i\int d\omega K_1^* e^{i\omega(t-t_0)} a_1^+ (\omega,t_0)\sigma_z (t)\cr
&&-2\int_{t_0}^t dt' \left[\sigma_z(t')\sigma^+ (t)\int d\omega |K_2|^2 e^{i\omega (t-t')}\right.\cr
&&\left.-\sigma^+(t)\sigma_z (t') \int d\omega |K_2|^2 e^{-i\omega (t-t')}\right]\cr
&&+2i\int d\omega K_2^* e^{i\omega (t-t_0)} a_2^+ (\omega,t_0)\sigma^+ (t)\cr
&&+2i\sigma^+ (t)\int d\omega K_2 e^{-i\omega (t-t_0)} a_2 (\omega,t_0).
\label{peffectivebatheq}
\end{eqnarray}
We now make the so-called first Markov approximation~\cite{gardinerpra85}, neglecting the frequency dependences of the bath couplings $K_i$ to obtain
\begin{equation}
\frac{d\sigma_z}{dt}=-\frac{i}{\hbar}[\sigma_z,H]-\gamma_1 (\sigma_z +1) -2i\sqrt{\gamma_1} (e^{i\phi_1} \sigma^+  a_1^{\mathrm{in}}-e^{-i\phi_1}a_1^{\mathrm{in}+} \sigma^-)
\label{finalsigmazeq}
\end{equation}
and
\begin{eqnarray}
\frac{d\sigma^+}{dt}&=&-\frac{i}{\hbar}[\sigma^+,H]-(\gamma_1/2+\gamma_{\varphi})\sigma^+\cr
&& -i\sqrt{\gamma_1} e^{-i\phi_1} a_1^{\mathrm{in}+}\sigma_z +i\sqrt{2\gamma_{\varphi}}  (e^{-i\varphi} a_2^{\mathrm{in}+}\sigma^++e^{i\varphi}\sigma^+ a_2^{\mathrm{in}}),
\label{finalsigmapeq}
\end{eqnarray}
where we have reparametrized $K_1=\sqrt{\gamma_1/(2\pi)} e^{i\phi_1}$ in terms of the decay rate $\gamma_1$, phase $\phi_1$ and $K_2=\sqrt{\gamma_{\varphi}/(4\pi)} e^{i\varphi}$ in terms of the pure dephasing rate $\gamma_{\varphi}$ and phase $\varphi$. We have also used the identities $\sigma^+\sigma^- =\frac{1}{2}(\sigma_z +I)$ and $\sigma_z\sigma^+=-\sigma^+\sigma_z =\sigma^+$. The `in' bath operators are defined as $a_i^{\mathrm{in}}(t)=(2\pi)^{-1/2}\int d\omega e^{-i\omega (t-t_0)} a_i (\omega,t_0)$.

\end{document}